\documentclass[sigconf,natbib=true,anonymous=false]{acmart}
\usepackage{bm}
\usepackage{subfiles}
\usepackage{dirtree}
\usepackage{enumitem}
\AtBeginDocument{%
  \providecommand\BibTeX{{%
    \normalfont B\kern-0.5em{\scshape i\kern-0.25em b}\kern-0.8em\TeX}}}

\setcopyright{acmcopyright}
\copyrightyear{2023}
\acmYear{2023}
\acmDOI{xx.xx/xx.xx}

\acmConference[WSDM '23]{ACM International Conference on Web Search and Data Mining}{June 03--05, 2023}{Singapore}
\acmBooktitle{ACM International Conference on Web Search and Data Mining,
  June 03--05, 2023, Singapore}
\acmPrice{15.00}
\acmISBN{978-1-4503-XXXX-X/18/06}

\begin{document}

\title{Diverse legal case search}

\author{Ruizhe	Zhang}
\email{u@thusaac.com}
\affiliation{%
  \institution{Tsinghua University}
  \city{Beijing}
  \country{China}
}
\author{Qingyao	Ai}
\email{aiqingyao@gmail.com	}
\affiliation{%
  \institution{Tsinghua University}
  \city{Beijing}
  \country{China}
}
\author{Yueyue	Wu}
\email{wuyueyue1600@gmail.com}
\affiliation{%
  \institution{Tsinghua University}
  \city{Beijing}
  \country{China}
}
\author{Yixiao	Ma}
\email{mayx20@mails.tsinghua.edu.cn}
\affiliation{%
  \institution{Tsinghua University}
  \city{Beijing}
  \country{China}
}
\author{Yiqun	Liu}
\email{	yiqunliu@tsinghua.edu.cn}
\affiliation{%
  \institution{Tsinghua University}
  \city{Beijing}
  \country{China}
}


\begin{abstract}
In last decades, legal case search has received more and more attention. Legal practitioners need to work or enhance their efficiency by means of class case search. In the process of searching, legal practitioners often need the search results under several different causes of cases as reference. However, existing work tends to focus on the relevance of the judgments themselves, without considering the connection between the causes of action. Several well-established diversity search techniques already exist in open-field search efforts. However, these techniques do not take into account the specificity of legal search scenarios, e.g., the subtopic may not be independent of each other, but somehow connected. Therefore, we construct a diversity legal retrieval model. This model takes into account both diversity and relevance, and is well adapted to this scenario. At the same time, considering the lack of dataset with diversity labels, we constructed a diversity legal retrieval dataset and obtained labels by manual labeling. experiments confirmed that our model is effective.

\end{abstract}



\maketitle


\section{Introduction}

Legal retrieval is becoming increasingly important in recent years.\citeN{kuhlthau2001information,makri2008investigating,russell2018information} 
Given a specific case (namely a query case), searching and analyzing similar cases from a large-scale legal document collection is a common practice for legal practitioners. \cite{bench2012history}
A high quality legal case retrieval system can assist users in effective and efficient analysis of the query case, and thus is of great value to modern legal systems.\cite{hamann2019german}


Existing studies on legal case retrieval mostly focus on how to rank and evaluate legal documents according to their individual relevance to the query. \cite{van2017concept}
Similar to classic IR problems such as ad-hoc retrieval and Web search, relevant documents in legal case retrieval usually are lexical or semantically similar to the query.
Therefore, early approaches often adapt retrieval models from Web search directly to legal case retrieval.\cite{koniaris2016multi}
For example, \citet{wen2006york} applied BM25\cite{robertson1994some} to retrieve relevant documents in legal case retrieval.
Advanced neural and language modeling approaches such as BERT\cite{devlin2018bert} and BERT-PLI\cite{shao2020bert} have also been used in legal case search. 
Experiments from previous studies have shown that these methods can achieve the state-of-the-art performance on retrieving relevant and similar documents on large-scale legal case retrieval benchmarks.


In practice, however, legal practitioners need more than similar documents. 
To make informed decision on a case, legal practitioners usually need to check about cases that not only share similar content with the current case, but also cover a variety of possible subtopics\footnote{We use charges as subtopics in our work. In the rest of this paper, subtopic(s) and charge(s) have the same meaning.} related to the case. 
Previous studies on legal case retrieval have already identified such needs of diversity~\cite{koniaris2017evaluation}.
Therefore, optimizing search diversity in legal case retrieval systems is important for improving system quality and user satisfaction.

\citet{koniaris2017evaluation}tried to adapt methods from open-domain search (e.g., Web search) directly to legal case retrieval, we believe that such paradigms could be suboptimal for multiple reasons.
First, the reason behind diversified search intents is different in legal case retrieval and open-domain search.
Generally, queries in open-domain search have shorter query words than those in legal search. Shorter query words have less information and more likely to be ambiguous.
Without additional information on the query intent, the best we could do is to diversify the search results to satisfy more users in open-domain search.
Legal case retrieval, on the other hand, usually has long and precise query in which users provide extensive details about the objective case. In legal case retrieval, users give the description of the case (with 7.07 setences in average, see Sec.2 for detail) as query word.
However, according our static result, users often want to explore multiple types of cases with different but related charges to justify the final judgment decisions. This is due to the characteristics of the  legal case retrieval task.  
Therefore, purely based on document-query similarity without considering the user's exploration need could be suboptimal for legal search in practice.

Second, diversification in legal case retrieval has unique challenges and opportunities. 
On the one hand, the restricted domain of legal case retrieval limits search diversification to focus on a limited number of query subtopics (i.e., charges) in the data collection. 
On the other hand, in contrast to open-domain search where query subtopics are often independent of each other, subtopics in legal case retrieval sessions often have correlations.
For example, when the query is relevant to `abandonment` , the user may also have the information needed on `abuse`. Therefore, the logical connection between charges can be important information in legal case retrieval.
Without mining the relationships between subtopics, open-domain diversity models~\cite{rodrygo2015search,carbonell1998use,agrawal2009diversifying,kharazmi2014using} that treat query subtopics could produce inferior performance in legal case retrieval.


In this paper, we study in depth on search diversification for legal case retrieval.
First, we 
build a new legal case retrieval dataset with a focus on search diversity. The language of the dataset is in Chinese. And all of 106 cases in the dataset are criminal cases.
In contrast to previous studies~\cite{koniaris2017evaluation} that evaluate search diversity with pseudo aspect relevance labels constructed with the latent topics (extracted by topic modeling approaches) of documents, our dataset is the first legal case retrieval dataset that contains explicit human annotations on query aspects (i.e., charges) and aspect-level document relevance judgments. 
Further, based on our observations in the user studies, we propose a search diversification algorithm specifically tailored for legal case retrieval.
We refer to it as the Diversified Legal case Retrieval Model (DLRM).
Instead of modeling search diversity purely based on the dissimilarities between documents, DLRM explicitly models subtopic relationships with a legal knowledge graph. 
The final results are ranked based on both the text similarity of query-document pairs and the relationships between diverse charges in legal case retrieval.
Our experiments show that DLRM can significantly outperform both non-diversity baselines and the state-of-the-art search diversification methods in legal case retrieval.

To summarize, the main contributions of this work are as follows: 
\begin{itemize}
    \item We build the first legal diversification dataset with human labels in both query-subtopic and subtopic-level relevance.
    \item We propose a Diversified Legal case Retrieval Model (DLRM) that improves the quality of legal case retrieval based on the interrelationships between queries and related charges.
\end{itemize}

The rest of the paper is organized as follows.
In Section 2, we describe the methodology of data collection and provide some basic information about the dataset.
We formally introduce DLRM in Section 3 and compare it with state-of-the-art diversification algorithms in experiments in Section 4. 
We review previous related work is Section 5 and discuss the conclusion and future work of this paper in Section 6.

\section{related work}
We briefly summarize related work in two categories: legal case retrieval and diversification. The former includes the mainstream methods in legal case retrieval. The latter reviews the diversity methods for web search.
\subsection{Legal case retrieval}

In recent years, more and more legal judgement are stores with digitization. Legal case retrieval becomes an important research issue for both IR and legal.~\cite{tran2020encoded,van2017concept,shao2020towards} Because of this,some approaches adapt retrieval models from Web search directly to legal case retrieval. And some legal case retrieval models also have been proposed~\cite{shao2020bert}. 

Different from web retrieval, users often input the whole describe of cases with long text to find similar cases. In web search, there are a series of ranking methods like BM25~\cite{robertson1994some}, TF-IDF~\cite{salton1988term}, LTR(Learning To Rank)~\cite{liu2011learning}. Deep learning methods like DSSM~\cite{hu2014convolutional}, CNN~\cite{shen2014latent}, RNN~\cite{pang2016text} and Match-SRNN~\cite{wan2016match} are also used to optimize the perform of ranking. However, unlike the case of short query terms in web retrieval, these well-know methods often do not perform well when the input is long query cases. When the query cases is long, we can analysis much more information about user requirements. We can take these information into account to optimize the ranking method. 

In legal case retrieval field, ~\cite{van2017concept} analysis the definition of relevance in law. And ~\cite{bench2012history} proposed a variety of approaches to legal case retrieval. ~\cite{shao2020bert} suggested BERT-PLI to improve legal case retrieval. But these methods did not consider the influence of charges while charge(s) of cases is an important reason to decide weather the result is relevant or not. 

\subsection{Diversification}

It is a common practice to construction a diverse ranking list is web search. Sometimes, users input the query text with ambiguity and redundancy. In order to satisfied users information requirement with diverse intents, a lot of methods are proposed~\cite{rodrygo2015search}.  \citet{carbonell1998use} proposed MMR(maximal marginal relevance) method to construct a novelty ranking list. RM(risk minimisation)~\cite{zhai2006risk} CR(conditional relevance)~\cite{chen2006less} MVA(mean-variance analysis)~\cite{wang2009portfolio} QPRP(quantum probability ranking principle)~\cite{zuccon2010using} ARW(absorbing random walk)~\cite{zhu2007improving} and SSSD(sparse spatial selection diversification)~\cite{gil2011sparse} were also proposed as a different way to optimize the novelty of result list. Coverage-based approaches are also useful to improve diversification of SERPs. RAB(Ranked-armed bandits)~\cite{radlinski2008learning} FM(facet modelling)~\cite{carterette2009probabilistic} and SD(score difference)~\cite{kharazmi2014using} are proposed to construct a list which coverage users intents. As Hybrid of novelty and coverage methods, WWC(weighted word coverage)~\cite{tsochantaridis2005large} DP(diversification perceptron)~\cite{raman2012online} RLTR(relational learning to rank)~\cite{zhu2014learning} DDF(diversified data fusion)~\cite{liang2014fusion} and IA-select~\cite{agrawal2009diversifying} was proposed. Then some diversification algorithm based on reinforcement learning(e.g. M2DIV\cite{M2DIV}) was proposed and state-of-the-art. These methods consider both novelty and coverage to improve the diversification of ranking list.

However, diversification for legal case retrieval is exactly different with web retrieval. First, as users often input a whole legal case, queries of legal retrieval have less ambiguity but more redundancy. Second, when the query case is referring to a certain charge, unlikely web search, user can also interested to cases referring to other charges. Users often have requirement to comparing, reference, discernment and contrast.
\section{Log Analysis on legal case retrival}
In this section, we present a log analysis to illustrate user's diverse intents in the legal case retrieval.
The data we analyzed are from a commercial legal case search engine.
In this search engine, users could submit keywords concerned in possible charges of candidate cases to the search box, and the system returns a list of cases highly related to the input charges.
We collected the real search logs of the search engine for 7 days and get 1905 search sessions in total.
Among them, 281 sessions were searching for criminal cases.
In this paper, we only focus on the criminal cases and treat the types of charges as the subtopics for queries.
We use ``subtopic'' and ``charge'' interchangeably in the rest of this paper. 


Previous studies on Web search diversification~\cite{guo2009efficient,agrawal2009diversifying} indicate that
\begin{itemize}
	\item If a query contains a certain word/phrase, the user's intent is likely to be related to the word/phrase.
	\item If a user clicks on a result related to a certain word/phrase, their intent is likely to be related to the word/phrase.
\end{itemize}
Similarly, in legal case retrieval, we assume that
\begin{itemize}
    \item If a query contains a certain charge, the user's intent is related to this charge.
    \item If a user clicks on a result with a certain charge, their intent is related to this charge.
\end{itemize}

\begin{figure}[t]
  \centering
  \includegraphics[width=\linewidth]{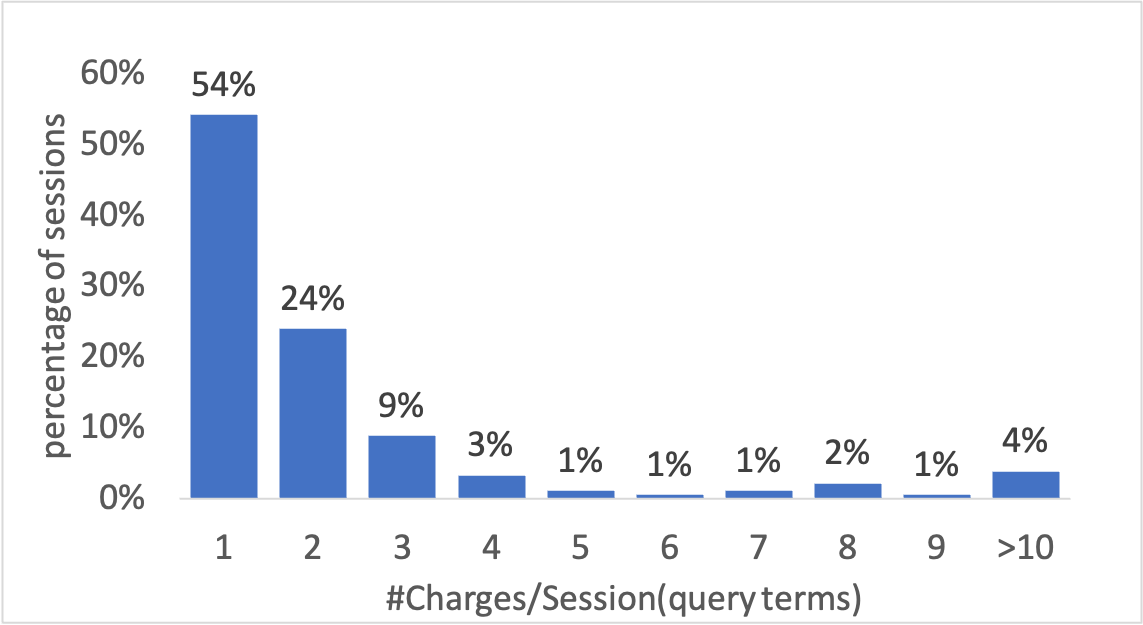}
  \caption{The graph shows the distribution of "the number of Sessions with certain number of charge(s) (in query terms)" ($\#Charges/Session\left(query\ terms\right)$ in short). We found that in more than $45\%$ sessions, users input more than one charges in query text. This shows that about half users describes they intents on diversity subtopics directly.}
  \Description{relevance_level}
  \label{fig:3RQ1}
\end{figure}

Based on the above premise, the first question we seek to answer is \textbf{RQ1:} \textit{Do users search for queries with different charges in one search session?} 
In legal case retrieval, users often submit multiple queries in a single search session.\cite{shao2020towards} 
Our goal here is to figure out whether users have diverse intents within a single session.
Thus, we count the number of sessions with a certain number different charges in the queries (\#\textit{Charges in queries per Session}) and show the results in Figure~\ref{fig:3RQ1}. 
As shown in the figure, $46\%$ of sessions in our search log have more than one charges in user queries directly. 
Also, over $20\%$ sessions involves at least 3 charges in the queries.
Therefore, it is reasonable to conclude that a noticeable number of users would directly search for queries with multiple charges in a single search session.

\begin{figure}[ht]
  \centering
  \includegraphics[width=\linewidth]{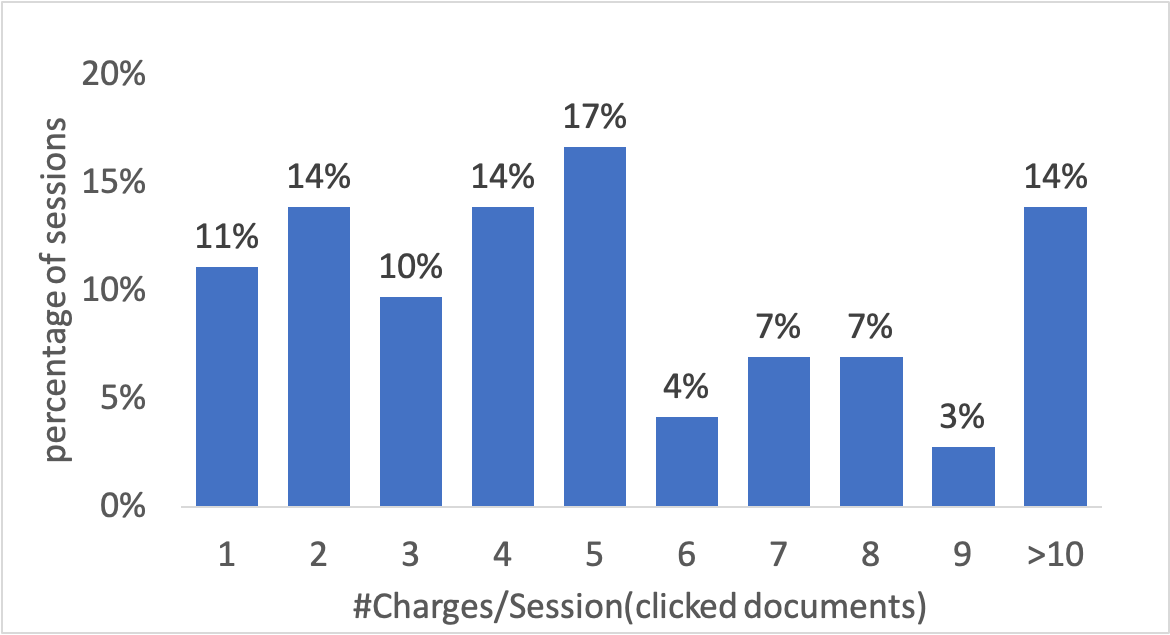}
  \caption{The graph shows the distribution of "the number of Sessions with certain number of charge(s) per session(involved by clicked documents)" ($\#Charges/Session\left(clicked\ documents\right)$ in short). We found that in more than $89\%$ sessions, the number of charges included by documents users clicked is more than one. Almost all users' click behavior reflects the diversity of their search intent.}
  \Description{relevance_level}
  \label{fig:3RQ2-1}
\end{figure}

The second question we want to investigate is
\textbf{RQ2:} \textit{Do users click on the results with different charges in a single legal case search session?}
Previous user studies in Web search~\cite{agrawal2009diversifying,chen2012beyond} show that users may not always indicate their search intents in their queries. 
In these cases, clicked documents in search can serve as an important signal to reflect user's intents. 
Therefore, we count the number of distinct charges included in the clicked documents in each session, which we refer to as \#\textit{Charges in clicked documents per Session}.
The results are shown in Figure~\ref{fig:3RQ2-1}
As depicted in the figure, $89\%$ of sessions have clicked documents that cover diverse charges. 
More than half of the users clicked on documents with as many as five charges in a single session.
This indicates that users in legal case retrieval often want to see results that cover a large number of charges in practice.

\begin{figure}[t]
  \centering
  \includegraphics[width=\linewidth]{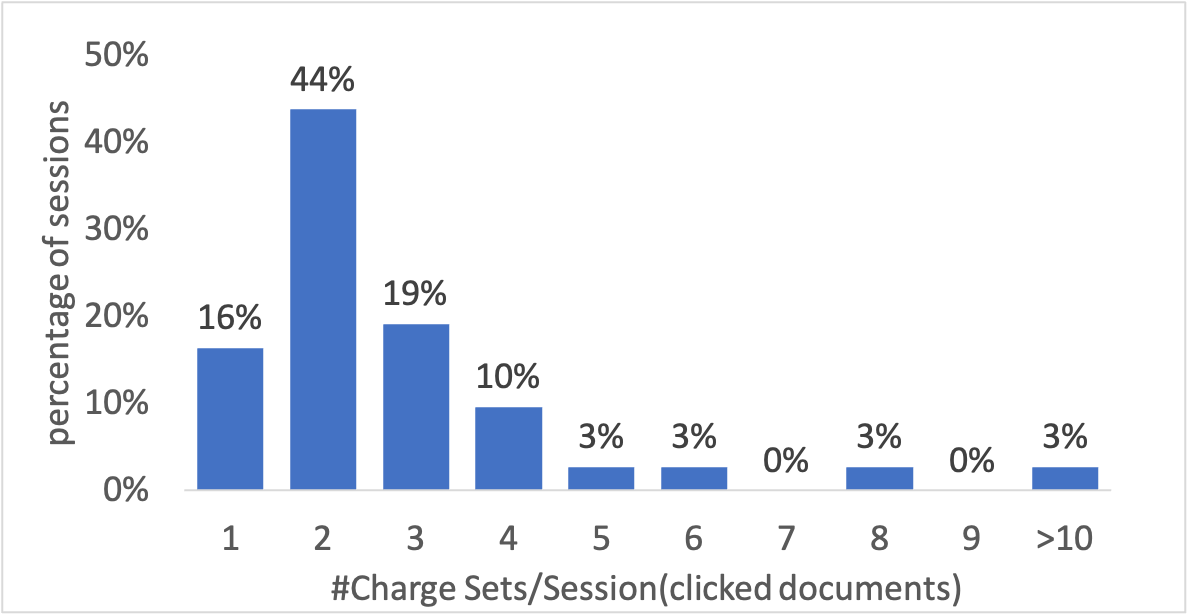}
  \caption{The graph shows the distribution of "the number of Sessions with certain number of Charge Sets  per session(included by clicked results)" ($\#Charge Sets/Session\left(clicked\ documents\right)$ in short). We found that in $84\%$ sessions, the number of charge sets included by documents users clicked is more than one. Most users' click behavior reflects the diversity of their search intent.}
  \Description{relevance_level}
  \label{fig:3RQ2-2}
\end{figure}

One potential flaw of counting distinct charges in the clicked documents is ignoring the fact that, in legal case retrieval, one document could have multiple charges directly.
It is risky to conclude that users have diverse intents when all clicked documents are covering the exact same set of charges.
Thus, to address this problem, we further count the distinct charge sets in the clicked documents in each session. 
For example, suppose that the user has clicked two documents $X,Y$, where $X$ covers charges $\{A,B\}$ and $Y$ covers charges $\{B,C\}$.
Then the number of distinct charge sets is considered as 2 ($\{A,B\}$ and $\{B,C\}$) here\footnote{We consider the subset of a specific set as a different set.}. 
We plot the distribution of \#\textit{Charge sets in clicked documents per Session} in Figure~\ref{fig:3RQ2-2}.
Our observations on Figure~\ref{fig:3RQ2-1} and Figure~\ref{fig:3RQ2-2} are similar.
About 84\% sessions have diverse user intents, which means that the users need documents that cover multiple charges. 
The percentage of users that need more than 3 distinct charge sets is around 22\%, and the number for users with 2 distinct charge sets is 44\%.
We can conclude that the majority of legal case retrieval users want to check about diverse search results in practice.

Considering the special properties of legal retrieval, the need of search diversification is in fact not surprising.
In contrast to Web search, legal search users often need to conduct search in depth. 
Legal practitioners often need to dig into different types of cases and charges in order to justify their decisions, and this has been reflected in their daily search behaviors.
Therefore, the importance of search diversity in legal retrieval is significant.
This motivates us to further build datasets and algorithms for legal search diversification.

\section{Dataset Construction}\label{sec:dataset}
In this section, we describe our lab study and the construction of a new legal case retrieval dataset with a focus on search diversity.
We refer to the dataset as Diversity Legal case Retrival Dataset (DLR-dataset).

\subsection{Overview}

Finding reliable and reusable datasets is an important step towards building effective retrieval models.
Multiple datasets~\cite{1997The,2009Overview} have been proposed for Web search diversification and lead to the development of a variety of successful algorithms.  
In the domain of legal retrieval, however, there isn't any public dataset for search diversification.
Previous studies propose to extend existing legal search datasets with pseudo aspect-level relevance labels by treating the latent topics (extracted by topic models) of each document as the subtopics of each query.
Unfortunately, subtopics extracted in this way are neither reusable nor reliable because the outputs of latent topic modeling approaches are usually unstable.
As a result, experiments on such datasets are difficult to be reproduced.

To avoid these problems, we build the DLR-dataset directly with human annotations.
In general, building a dataset for search diversification involves two parts: the identification of query subtopics, and the annotation of subtopic-level document relevance.
Thus, the main goal of our lab study contains the following two parts:
\begin{itemize}
    \item Understanding the distribution of user intents (on charge levels) in a certain query case.
    \item Understanding if a candidate document can satisfied a user's search intent on a specific charge.
\end{itemize}
Our lab study is conducted based on a legal case retrieval dataset for Chinese law system~\cite{ma2021lecard}. The language of this dataset is Chinese. This dataset includes 107 criminal cases. And under each case, 100 judgments were provided as candidate documents.
We use 106 query cases in this dataset (removed one query case because it includes 29 sentences, which is much longer than others). 
For these 106 query cases, we measured the number of sentences in each query case. Result in table \ref{sen_num} to show that the query terms in legal case retrieval are indeed much longer than in web search.

\begin{table}[ht]
  \caption{Number of sentences in each query case. Most of query cases includes 5 to 10 sentences. The shortest one has 2 sentences while the longest one has 20 sentences. And 7.07 sentences/query case in average. It is much longer than that on web search(often only piece of words, no sentences). }
  \Description{Number of sentences in each query case.}
    \begin{tabular}{lccccc}
		\toprule
		\#Seteneces &$\leq 5$&$(5,10]$&$(10,15]$&$>15$\\
		\midrule
		Percentages & $35.85\%$ &$39.62\%$ &$19.81\%$&$5.66\%$\\
		\bottomrule
	\end{tabular}
	
	\label{sen_num}
\end{table}

For every query case, we use the top 30 candidate documents retrieved by BM25 in this dataset. For detail, we first use THULAC(THU Lexical Analyzer for Chinese) \cite{THULAC} to divide words. And then use BM25 to calculate the relevance score between query case and each candidate case.
Specifically, we construct the DLR-dataset with two steps.
First, we annotate the relevance between queries and possible charges. 
Second, we label the relevance of each query-charge-document triple based on the results from the first step.
More details about the specific notations used in this paper are shown in Table~\ref{table4_1}.
To be specific, we define variables below:
\begin{table}[t]
  \caption{Definitions of Notations}
  \begin{tabular}{ll}
    \toprule
    Variable & Description\\
    \midrule
    $n=107$ & \#query cases\\
    $m=30$ & \#candidate cases/query cases\\
    $s=272$ & \#charges\\
    $Q=\{q_i\}\left(i\in [1,n]\right)$ & query cases\\
    $I=\{I_{k}\}\left(k\in [1,s]\right)$ & intents on charges\\
    $D=\{d_{ij}\}\left(i\in [1,n],j\in [1,m]\right)$ & candidate documents\\
    \bottomrule
  \end{tabular}
  
  \label{table4_1}
\end{table}


\subsection{Query-charge Relevance Annotation}

In the first step, our goal is to identify possible user intents on charges (i.e., $\{I_k\}$) in a specific query case $Q_i$. 
In particular, we are interested in understanding the distribution of needs for different charges in users who submitted the query.

\subsubsection{Annotation Process}\label{sec:query_charge_annotation_process}
Annotating all possible query-charge pairs is prohibitive due to the large number of queries and charges in the dataset.
Instead, we adopt a two-step process to create the candidate charge pool for each query.
First, we use regular expressions to extract charges in the query string.
Second, we use a legal judgement prediction (LJP) model\cite{zhong2018topjudge} to predict top-5 relevant charges to the query case, and merge them with the extracted charges from the first step to form the final candidate charge set (CCS).

In the actual annotation process, we recruited 8 annotators to label the relevance between queries and charges.
All annotators are legal practitioners with sufficient knowledge backgrounds in law.
Specifically, we asked each annotator to first read the descriptions of the query cases and CCS, and then select and sort candidate chages for each query.
For a specific query, the annotator were instructed to first choose the relevance charges from the query's CCS, and then sort them according to how important/relevant the charges are to the query.
For example, we may give a CCS$=\{I_1,I_2,I_3,I_4,I_5,I_6\}$, and the output of an annotator could be a sorted list such as $I_2=I_3>I_1>I_5=I_6$ where $I_4$ is ignored since the annotator thinks it is not relevant.
Also, since the CCS we created can hardly cover all relevant charges for each query, we asked the annotators to submit new charge(s) that they think are relevant to the query case. 
In our study, however, none of the annotators have submitted new charge to the dataset. 
This indicates that the method we used to collect charge candidates is effective.


\subsubsection{Result Analysis}
Here we briefly discuss the results we get from the query-charge relevance annotation.
First, we show the number of distinct relevant charges we get for each query in Table~\ref{fig:intent size}.
We simply merge the annotated relevant charges from all annotators to form the final charge set of each query.
As shown in the figure, all queries in our dataset have at least 2 relevant charges.
Also, in excess of 90\% of the queries have more than 3 relevant charges.
This indicates that legal search users often have strong needs to check about documents referring to multiple relevant charges.

\begin{table}[ht]
\caption{Through the statistics of the labeling results, we can get the following conclusions. 1.In all of 106 queries, participants have diversity intents for a certain query case. 2.In most of cases, size of intents user may have are between 3 to 5.}
  \begin{tabular}{lcccccc}
        \toprule
       Size of intent(s) set & 1& 2&3&4&5&6\\
       \midrule
       Percentage & $0.0\%$&$6.5\%$&$35.5\%$&$33.6\%$&$21.5\%$&$2.8\%$\\
       \bottomrule
  \end{tabular}

  \Description{size of intents}
  \label{fig:intent size}
\end{table}

Second, we analyze how the importance of each relevant charge distribute in each query.
With the select-and-sort annotation process, we collected annotators' preferences over relevant charges in each query, which can then be used to create multi-level relevance labels for charges.
Table~\ref{fig:relevance level} shows how many relevance levels the annotators have created for each query.
For example, a sorted list $I_2=I_3>I_1>I_5=I_6$ selected from a CCS$=\{I_1,I_2,I_3,I_4,I_5,I_6\}$ means 4 relevance level for charges, e.g., \textit{perfect} for $\{I_2, I_3\}$, \textit{excellent} for $\{I_1\}$, \textit{fair} for $\{I_5, I_6\}$, and \textit{irrelevant} for $\{I_4\}$.
\begin{table}[ht]
  \caption{Levels of importance(LoI in short) participants divided. In $48.1\%$ cases, participants only divide results into 2 LoI while in $41.9\%$, participants divide into 3 LOI. Less than $1\%$ participants, user divide intents into more than 5 LoI.}
  \Description{relevance_level}
    \begin{tabular}{lccccc}
		\toprule
		LoI &2&3&4&5&6\\
		\midrule
		percentage & $48.1\%$&$41.9\%$&$8.9\%$&$0.9\%$&$0.1\%$\\
		\bottomrule
	\end{tabular}
	
	\label{fig:relevance level}
\end{table}

As shown in the table\ref{fig:relevance level}, 48.1\% of the queries only have two-level relevance judgments, i.e., \textit{relevant} and \textit{irrelevant}, 41.9\% of the queries have three-level judgments, and about 10\% of the queries have more than three levels.
In other words, the number of relevance levels in the annotation results vary significantly among different queries. 
This indicates that using a relevance grading method with a fixed number of possible levels is not suitable for query-charge annotation.

Other than the importance of each charge, legal system designers may care more about the distribution of user intents in each query, namely $P(I_k|Q_i)$.
Unfortunately, there is no simple solution to get such information without large-scale user survey in practice.
In this paper, we adopt a naive strategy to compute intent distributions based on the annotated query-charge pairs.
First, for each sorted intent list, if the annotator divided results into $k$ levels, we choice $k$ value(s) in range $[0,1]$ uniformly to represent the probability of having each intent in the query. 
For example, if a sorting list is $I_2=I_3>I_1>I_5=I_6$ with $I_4$ unselected, the probability of having each charge as the query intent is computed as 1 for $\{I_2,I_3\}$, $\frac{2}{3}$ for $\{I_1\}$, $\frac{1}{3}$ for $\{I_5, I_6\}$, and 0 for $\{I_4\}$.
Then, we average the intent distribution collected from all annotators to get the final $P(I_k|Q_i)$.





\subsection{Charge-level Relevance Annotation for Query-document Pairs}

Given the relevance annotation on query-charge pairs, in the second part of our lab study, we want to collect fine-grained relevance information for each query-charge-document triples $(Q_i, I_k, d_{i,j})$.

\subsubsection{Annotation Process}\label{sec:doc_annotation}

In the actual annotation process, we recruited 9 annotators(include previous 8 annotators) to label the relevance between queries and charges.
All 9 annotators are legal practitioners with sufficient knowledge backgrounds in law.

We divided the 9 annotators into 3 groups evenly and randomly. 
We asked each group to annotate the documents in 35 (or 36) queries. We collect all labels from 3 annotators and record them in the dataset. And we use the median of scores in each group to be the final relevance labels for the $(Q_i, I_k, d_{i,j})$ triples in following experiments. 

We found that the amount of work involved in annotating all the triples $(Q_i, I_k, d_{i,j})$ would be enormous and unacceptable. The numbers of $Q$, $I$ and $d_{i,j}$ are $106$, $272$ and $100$. Annotating a triple may cost an annotator $2~4$ minutes($3$ minutes in average). The total time cost will be $106\cdot 272 \cdot 100 \cdot 3=8,649,600$ minutes, which is about 16 years. This is unacceptable at all.
And we believe that a candidate judgement `may` satisfied the users' information need of intent $I_k$, only if the user has information need of intent $I_k$ when search the query case $Q_i$ and the document $d_{i,j}$ is relevant to the intent $I_k$.

So , in this step, we only let annotators to label part of triples.
A triple $(Q_i, I_k, d_{i,j})$ would be labeled by annotators if and only if all three conditions are meet: 
\begin{itemize}
    \item (1) the probability of observing the charge in the user intent of the query ($P(I_k|Q_i)$) is above zero (labeled in step 1);
    \item (2) the document $d_{i,j}$ is relevant to the query $Q_i$(labels from the LeCard which we based on\cite{ma2021lecard}); 
    \item (3) the document $d_{i,j}$ is relevant to the charge $I_k$(filter from 272 subtopics).
\end{itemize}

To filter charges in condition 3, we do the same filter as in step 1. We merge the LJP modles' top-5 relevant charges and charges extracted by regular expressions as relevant charges of document $d_{i,j}$.

In this way, the workload of the annotator is greatly reduced. Statistically, an average of  $1858.3$ triples need to be annotated per annotator. As a result of this improvement, each annotator spent approximately $1858.3\cdot3=5574.9$ minutes (about $92$ hours) amount of time on this step of the annotation process.

A triple $(Q_i, I_k, d_{i,j})$ would have a non-zero relevance label if and only if: (1) the probability of observing the charge in the user intent of the query ($P(I_k|Q_i)$) is above zero; (2) the document $d_{i,j}$ is relevant to the query $Q_i$; and (3) the document $d_{i,j}$ is relevant to the charge $I_k$.
In this paper, the documents we consider are legal cases with judgments, which means that we can directly extract the relevant charges of a document from its content.
After extracting the relevant charges of each document, we recruited 9 annotators (again, all annotators are legal practitioners with sufficient knowledgeable background in law) and let them assess a four-level relevance label (i.e., \textit{perfect}, \textit{excellent}, \textit{fair}, and \textit{irrelevant}) for each triple $(Q_i, I_k, d_{i,j})$.







\begin{table*}[ht]
	\caption{Fleiss kappa and Kendall rank correlation coefficient. We divide participants into 3 groups(A,B,C), 3 participants in the same group label on same tasks. We calulate kappa and Kendall rank for every pair of users in same group. It shows that there is a good agreement between participants. The quality of label results is excellent.}
	\begin{tabular}{cc|cccc|cccc|cccc}
		\toprule
		Measure&Group&&&A&&&&B&&&&C&\\
		\midrule
		&User1&&A1&A2&A3&&B1&B2&B3&&C1&C2&C3\\
		\midrule
		&User2&A1&-&0.49&0.42&B1&-&0.4&0.59&C1&-&0.47&0.38\\
		Kappa&User2&A2&0.49&-&0.4&B2&0.4&-&0.55&C2&0.47&-&0.45\\
		&User2&A3&0.42&0.4&-&B3&0.59&0.55&-&C3&0.38&0.45&-\\
		\midrule
		&User2&A1&-&0.89&0.67&B1&-&0.72&0.72&C1&-&0.74&0.80\\
		Kendall's $\tau$&User2&A2&0.89&-&0.56&B2&0.72&-&0.86&C2&0.74&-&0.70\\
		&User2&A3&0.67&0.56&-&B3&0.72&0.86&-&C3&0.80&0.70&-\\
		\bottomrule
	\end{tabular}
	
	\label{quality}
\end{table*}

\subsubsection{Result Analysis}
We compute and visualize the Fleiss kappa and Kendall rank correlation coefficient to evaluate the quality of label results in Table~\ref{quality}.
Overall, the agreements between annotators are high and the labels we get from the annotation process is reliable. 
Further, we randomly split the queries into a training set and a test set with ratio 2:1. 
The final statistics of the dataset is shown in Table~\ref{dataset_basic}.
The dataset will be shared publicly to the research community when this paper is published.
\begin{table}[t]
  \caption{The basic information of the diversity legal case retrieval dataset.}
  \begin{tabular}{ccc}
    \toprule
     & Training & Test\\
    \midrule
    \#Querys &70 & 36\\
    \#Candidate / Query & 30 &30\\
    \#Queries' Relevant charge(s) & 3.54 &3.50\\
    \#Sentences / Query(Avg.)  & 7.78 & 7.56\\
    \#Sentences / Document(Avg.)  & 188.12 &181.69\\
    \midrule
    \bottomrule
  \end{tabular}
  
  \label{dataset_basic}
\end{table}





\section{Diversified Legal Case Retrieval Model}
\begin{figure}[h]
  \centering
  \includegraphics[width=1.0\linewidth]{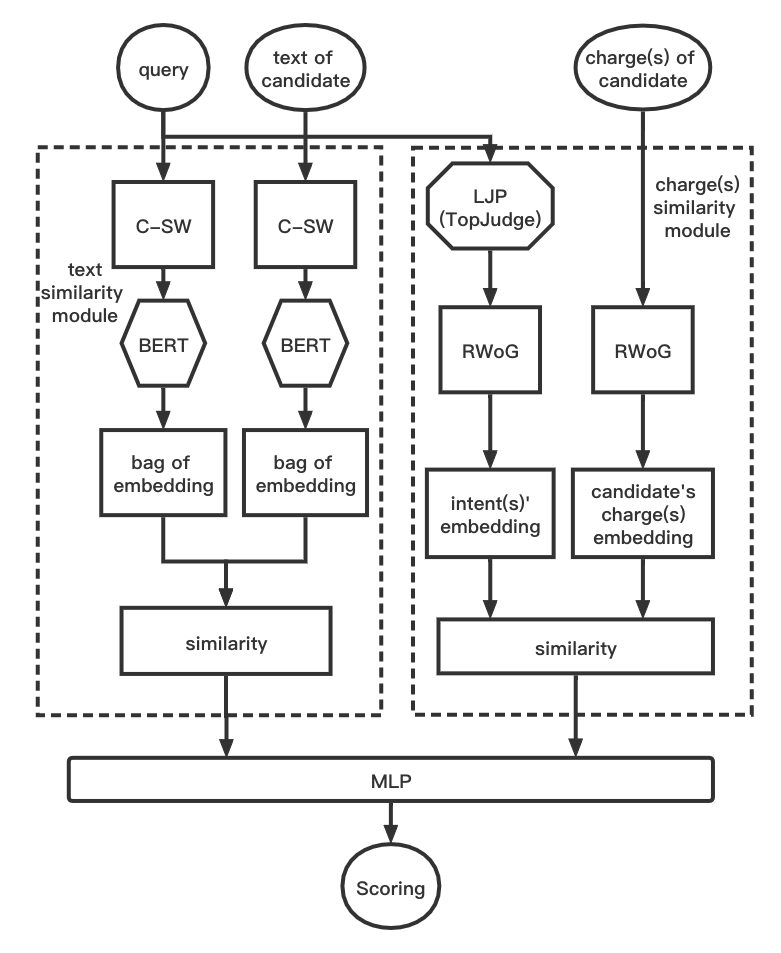}
  \caption{An overview of the Diversified Legal case Retrieval Model (DLRM). The DLRM contains a text similarity module and a charge(s) similarity module, and the output of these modules are combined with a MLP model. `C-SW` in text similarity module means Cut with Sliding Windows. The `RWoG` in charge(s) similarity module refer to the Random Walk on Graph process. }
  \Description{relevance_level}
  \label{fig:module_en1}
\end{figure}
In this section, we introduce the Diversified Legal Case Retrieval Model (DLRM). 
The DLRM consists with a text similarity module, a charge similarity module, and an multi-layer perceptron model (MLP) that combines the output of both modules. 
The overall structure is shown in Figure~\ref{fig:module_en1}.
Specifically, the inputs of the modules are a query case $Q_i$ and a candidate document $d_{ij}$. 
For each document, we also have several charges associated with it (as described in Section~\ref{sec:doc_annotation}). 
The final output of the DLRM is the ranking score of the candidate document. 

\subsection{Text Similarity Module}


Text similarity is an important signal for document relevance.
Thus, in DLRM, we explicitly design a text similarity module to capture the lexical and semantic similarities between query and document text.
Specifically, the input of the text similarity module is the raw text of a query case and a candidate document. 
The output of the module is an embedding $T_{sim}$ that encodes the semantic information extracted from the raw text.

Legal documents are usually too long to be processed by neural language models directly.
To address this problem, in this paper, we adapt a sliding window to cut long text into small passages with overlaps, which we refer to as the Cut with Sliding Windows (C-SW) module.
Formally, let the input of the C-SW module is a piece of text with $l$ sentences $\{s_1,s_2,...,s_l\}$.
Given a sliding window with size $w$ and step $d$, the first output passage of the C-SW module would be $\{s_1,s_2...s_w\}$ ,and the second would be $\{s_{1+d},s_2...s_{w+d}\}$, etc. 
We pad the output passage with empty strings to ensure that each passage have $w$ sentences.
Specifically, we set $w$ and $d$ to be 3 and 1 for an input query, and 13 and 5 for an input document, respectively.

For each output passage from the C-SW module, we use a pre-trained BERT model to encode and build an embedding representation of the passage (i.e., the 768 dimensional vector of [CLS] in BERT).
Let $n$ and $m$ be the number of passages we extracted for a query and a document, respectively.
We can compute a similarity matrix $M\in \mathbb{R}^{n\times m}$ where $M_{i,j}$ is the cos similarity of the $i$th passage of the query and the $j$th passage of the document.
Then, we apply a max-pooling layer over query passages (i.e., rows in $M$) to extract a similarity vector $T_s\in\mathbb{R}^{n}$.
Further, to create an input vector with a fixed length for the MLP model, we concatenate $\{T_s, \phi,T_s\}$ where $\phi$ is a sequence of 0s to form a vector $T_{sim}$ with a fixed length (e.g., 54 in our experiments).  



\subsection{Charge Similarity Module}\label{sec:charge_similarity}

As discussed in Section~\ref{sec:dataset}, legal search users often need documents that are relevant to multiple charges, and the number of possible charges in a legal corpus is limited.
In contrast to Web search, we observe strong correlations between documents with certain charges in our user study.
In other words, the relevance of query-charge-document triples $(Q_i, I_k,d_{i,j})$ with different charges are not independent to each other.
Without considering such information, existing search diversification algorithms could produce suboptmal results in legal case retrieval. 

To address this problem, we propose to build a charge-similarity module that encode the relationships of charges for legal case retrieval.
We first build a legal knowledge graph with charges based on the extracted charges of each query and document.
After that, we run a random walk algorithm, which we refer to as the Random Walk on the Graph (RWoG), to capture the semantic relationships between query charges and document charges.
Finally, we use the embedding of the query and the document on the graph to compute the charge similarity between them, which is then used as the output of the module.

Specifically, our legal knowledge graph is built with the judgments of different trials in cases.
In our dataset, most documents contain cases with only one trial and judgment, but there are indeed cases where multiple trials were conducted and the judgement of the first trial was reversed by the judgements of the later trails.
If the charge in a judgment is reversed by another charge in the later judgment, it is likely that the two charges have strong connections to each other. 
For example, they could be two charges that are difficult to be distinguished, which indicates that it could be helpful to show both charges to users in legal retrieval. 
Based on this observation, we build a legal charge graph with reverse information of charges.
First, we count the frequency of a charge $i$ being reversed by a charge $j$.
Let this frequency matrix be $G\in \mathbb{N}^{s\times s}$ where $s$ is the total number of possible charges in our dataset (i.e., $s=272$).
Then, we treat each charge as a node and build the directional edges among them in the following way:
\begin{itemize}
	\item For a node $i$, if $\forall $j$~G_{ij}=0$, then we add an edge from node $i$ to itself with weight 1.
	\item For a node $i$, if $\exists$j$~G_{ij}>0$, add an edge from $i$ to $j$ with weight $E_{ij}$ as 
\end{itemize}
\begin{equation}
	E_{ij}=
	\begin{cases}
		\alpha & \text{ if } i=j \\
		(1-\alpha)\cdot \frac{G_{ij}}{\sum_{k=1}^s G_{ik}} & \text{ if } i\neq j
	\end{cases}
\end{equation}

Based on the knowledge graph, we run a random walk on the graph (RWoG) to extract the charge-based embedding representations of the query and the document.
Let $C_{qo}\in \mathbb{R}^s$ and $C_{do}\in \mathbb{R}^s$ be binary vectors representing whether a charge is relevant to the query and the document, respectively.
For each query, we build $C_{qo}$ with the query's charge candidate set (as discussed in Section~\ref{sec:query_charge_annotation_process}).
The value of the $i$th dimension of $C_{qo}$ is set to be the output of the legal judgment prediction model, which can be treated as the probability of the query containing the charge.
For each document, we build $C_{do}$ with the document's charge candidate set (as discussed in Section~\ref{sec:doc_annotation})
The $i$th dimension of $C_{do}$ is 1 if the charge $i$ is relevant to the query (or document), and is 0 otherwise.
We normalize both $C_{qo}$ and $C_{do}$ to make the sum of their elements to be 1.
After that, we set the initial node probabilities with $C_{qo}$ and $C_{do}$ separately, and run RDoG for two times to extract new node probability vectors $C_{q}\in \mathbb{R}^s$ and $C_{d}\in \mathbb{R}^s$ as the final charge representations of the query and the document, respectively.

The final output of the charge similarity module, i.e., charge similarity embedding $C_{qd}$ between the query and the document, is computed as
\begin{equation}
    C_{qd} = C_{q}\bigotimes C_{d}
\end{equation}
where $C_{qd}$ is the Kronecker product of $C_{q}$ and $C_{d}$.

\subsection{Ranking Prediction and Model Training}\label{sec:training}

To create a final ranking for candidate documents, we concatenate the output of the text similarity module (i.e., $T_{qd}$) and the charge similarity module (i.e., $C_{qd}$) to form an input vector, and then use a multi-layer perceptron network (MLP) to predict the ranking score of each document.
We then sort documents according to their ranking scores to create the result list.
One thing to note is that, although we do not explicitly model document novelty in the ranking process, the charge similarity module of DLRM has already helped the model estimate ranking scores based on the distribution of the query intents.
Therefore, the results produced by our DLRM can capture search diversity implicitly. 
We further show the effectiveness of DLRM in search diversification in Section~\ref{sec:exp}.









Considering the limited size of our training data, given a target ranking metric (e.g., NDCG-IA@10), we adopt the following method to train our model.
First, we randomly choose a query $Q_i$ with corresponding candidate documents $\{d_i\}$ from the training set.
Second, for a particular document $d_{ij}$, we randomly choose a position $k$ (i.e., $k\in[1,10]$) and put $d_{ij}$ on $k$.
Then, we randomly fill other positions with documents randomly picked (without replacement) from $\{d_i\}$ and compute the expected metric rewards (e.g., NDCG-IA@10), which we refer to as $\mathbb{E}(R(k,d_{ij}))$, of the random sampled ranked list.
The final label $l(k, d_{ij})$ we assign to the document $d_{ij}$ is computed as
\begin{equation}
	l(k, d_{ij}) =\frac{\mathbb{E}(R(k,d_{ij}))-\min_a \mathbb{E}(R(k,d_{ia}))}{\max_a \mathbb{E}(R(k,d_{ia}))-\min_a \mathbb{E}(R(k,d_{ia})))}
\end{equation}
Let the ranking score of $d_{ij}$ be $\gamma_{ij}$ (i.e., the output of MLP).
We train our model by minimizing the mean square errors between $l(k, d_{ij})$ and $\gamma_{ij}$.
To make the whole training process more reliable, we repeat this process for 1 million times to produce the final model.

\section{Experiment}\label{sec:exp}

\begin{table*}[htbp]
	\caption{Experiment result on diversity legal case retrieval dataset. 
		$N-IA$ stands for $NDCG-IA$ and $\alpha-N$ stands for $\alpha$-NDCG. 
		* and ** denote significant differences with respect to the best baseline (exIA-select) at $p < 0.05$ and $p<0.01$ level using the pairwise t-test, respectively.}
\begin{tabular}{cllllllll}
	\toprule
	& $N-IA@1$ & $N-IA@3$ & $N-IA@5$ & $N-IA@10$ & $\alpha-N@1$ & $\alpha-N@3$ & $\alpha-N@5$ & $\alpha-N@10$\\
	\midrule
	BM25~ & $0.4537 $ & $0.4783 $ & $0.4921 $ & $0.5278 $ & $0.5448$ & $0.4970$ & $0.5085$ & $0.5621 $\\
	\midrule
	MMR & $0.4537 $ & $0.4769 $ & $0.4978 $ & $0.5181 $ & $0.5448$ & $0.5053$ & $0.5302$ & $0.5621 $\\
	IA-select & $0.4951 $ & $0.5070 $ & $0.5194 $ & $0.5548 $ & $0.5686$ & $0.4727$ & $0.4520$ & $0.4443 $ \\
	exIA-select & $0.6023 $ & $0.5971 $ & $0.6069 $ & $0.6370 $ & $0.7419$ & $0.6291$ & $0.6185$ & $0.6286 $ \\
	M2DIV & $0.5569 $ & $0.5505 $ & $0.5611 $ & $0.5778 $ & $0.6238$ & $0.5485$ & $0.5586$ & $0.5858 $\\
	\midrule
	DLRM & \bm{$0.7199^{**} $} & \bm{$0.7389^{**} $} & \bm{$0.7753^{**} $} & \bm{$0.8747^{**} $} & \bm{$0.8143^{*} $} & \bm{$0.6786^{*} $} & \bm{$0.6521^{*} $} & \bm{$0.6450 $}\\
	\midrule
	improve. & $19.5\%$ & $23.7\%$ & $27.7\%$ & $37.3\%$ & $9.8\%$ & $7.9\%$ & $5.4\%$ & $2.6\%$\\
	\bottomrule
\end{tabular}

\label{overall_result}
\end{table*}

In this section, we describe our experiments in details. 
We first introduce the experiment setup, and then discuss the overall results and ablation study of different diversification models.

\subsection{Experimental setup}

We conduct experiments with the proposed DLR-dataset described in Section~\ref{sec:dataset}, and we adopt two popular evaluation metrics for search diversification.
The first one is $\alpha$-NDCG.
The computation of $\alpha$-NDCG assumes that all search intents distributed evenly. 
To this end, we filter out charge $I_k$ with $P(I_k|Q_i)\leq0.5$ for each query $Q_i$ and use the rest as relevant intents for $Q_i$ in the computation of $\alpha$-NDCG.
Also, in $\alpha$-NDCG, each document can only be \textit{relevant} or \textit{irrelevant} to a query intent.
Thus, we convert the four-level relevance labels of each query-document-charge triple to a binary label (2,3 as 1, and 1,0 as 0) for the computation of $\alpha$-NDCG.
The second metric we used is NDCG-IA. 
Specifically, we use $P(I_k|Q_i)$ from the ground truths as the weight of each intent NDCG-IA: 
\begin{equation}
	\text{NDCG-IA}(Q_i) = \sum_{I_k} P(I_k|Q_i)\cdot NDCG(Q_i|I_k)
\end{equation}
where $NDCG(Q_i|I_k)$ is computed with the four-level relevance labels of each query-document-charge triple.

For comparison, we implement five baselines for our experiments:

\begin{itemize}
	\item \textbf{BM25}~\cite{robertson1994some}: A classic retrieval model which uses the BM25 function to measure the relevance between query cases and candidate cases. It doesn't consider search diversity.
	\item \textbf{MMR}~\cite{carbonell1998use}: A famous diversification algorithm that ranks documents based on both their relevance scores and novelty. Specifically, MMR computes the ranking score of a document as a linear combination of its relevance score to the query and its novelty compared to the previously selected documents in the list.
	\item \textbf{IA-select}~\cite{agrawal2009diversifying}: A state-of-the-art diversification algorithm that ranks documents based on their relevance to each query intent. Specifically, it computes the relevance scores of a document for each query intent separately and create the final ranking by balancing documents relevant to different query intents. Here we use the initial intent distribution extracted by LJP (i.e., $C_{qo}$) as the intents for each query. 
	\item \textbf{exIA-select}: A extended version of IA-select that uses the query intent vector learned by RWoG in the charge similarity module of DLRM (i.e., $C_q$) as the intent distributions of each query.
	\item \textbf{M2DIV}~\cite{M2DIV}: A state-of-the-art diversification algorithm that constructs a policy-value network with reinforcement learning to diversify search results. 
\end{itemize}

For the implementation of relevance models in MMR, IA-select, and exIA-select, we follow the experimental design proposed by \citet{devlin2018bert} and use a BERT model to encode both the query and the document into latent vectors. 
The relevance score of them is computed as the cos similarity between the BERT vectors of the query and the document.
For the MMR algorithm, we compute the novelty of a document as the averaged cosine similarity between it and the selected documents.
We tune the hyperparameter of the linear combination function from 0 to 0.1.
For other baselines, we use grid search to find the best hyper-parameters for them.
We only report the baseline performance with the best hyper-parameter settings we found.

For the DLRM, all parameters except those for the MLP network are fixed after the pre-train process. 
We only train the MLP network based on the training data. 
Specifically, we use the adam optimizer with loss@$10^{-5}$ to train MLP model. 
The size of the three hidden layers in the MLP is 128, 32, and 4. 
The $\alpha$ in RWoG is set to be 0.4.
The LJP module in our experiment is the TopJudge\cite{zhong2018topjudge}.
Also, to smooth the initial intent distribution predicted by LJP, we added $0.3$ to the LJP outputs of the top 5 charges of each query and then do the normalization as described in Section~\ref{sec:charge_similarity}.

\subsection{Overall Results}

Table \ref{overall_result} reports the performance of our DLRM and all baseline methods. 
As shown in the table, the DLRM has outperformed all baselines in our experiments.
Particularly, the improvement of the DLRM over the best existing search diversification baseline in our experiments (i.e., exIA-select) is 20\% or more on NDCG-IA.  
This demonstrates the effectiveness of DLRM as a search diversification model for legal case retrieval.

We found that DLRM outperforms the baseline model on all metrics including NDCG-IA@1. This shows that DLRM not only improves the quality of the ranking result list but also improves the quality of the top 1 results because it takes into account the diverse information need. 

To further illustrate the benefit of charge similarity modeling in legal case retrieval, we compare the results of IA-select with exIA-select where we replace the intent (charge) distribution (i.e., $C_{qo}$) used in IA-select with the one learned from our legal knowledge graph (i.e., $C_{q}$).
As we can see in Table \ref{overall_result}, exIA-select has outperformed IA-select on all metrics.
Their differences on all metrics are statistically significant.
This indicates that the charge similarity module of our DLRM can effectively extract the distribution of query intents and thus produces better search diversification models.

One interesting observation in our experiments is that the performance of MMR is quite similar to the performance of BM25. 
Considering the fact that MMR was implemented with a stronger model (i.e., BERT), this indicates that the incorporation of document novelty in MMR did not improve the overall performance of the algorithm.
In fact, we indeed observe that smaller weights of document novelty in MMR usually produce better results in our experiments.
This means that search diversification algorithms that only model diversity as the differences between documents could not satisfy the need of legal case retrieval users.


\subsection{Ablation study}

In this paper, we construct DLRM with two modules: the text similarity module and the charge similarity module.
To further show the effectiveness of each module, we further conduct a ablation study.
Specifically, we design three variations of the DLRM as the followings:
\begin{itemize}
	\item \textbf{Text Only}: The DLRM with the text similarity module and the MLP network only. 
	\item \textbf{Charge Only}: The DLRM with the charge similarity module and the MLP network only.  
	\item \textbf{None (Random)}: The DLRM with the MLP network only. In this case, the input of the MLP network is randomly initialized for each query-document pairs. 
\end{itemize}

Table~\ref{ablation_NDCG} shows the ranking performance of each model.
As shown in the table, all variations of the DLRM performed worse than the complete DLRM.
We observed that the charge-only model perform the best among all variations.
This indicates the importance of charge similarity modeling in legal case retrieval.

\begin{table}[htbp]
	\caption{An ablation study exploring that every part of DLRM is conducive to improve DLRM's performance on NDCG-IA@k. In this table,  
	The boldface highlights the best-performing setting among all settings. */** denotes that NDCG-IA@k performs significantly compared with complete model at $p < 0.05/0.01$ level using the pairwise t-test. N-IA is NDCG-IA in short.}
	\begin{tabular}{ccccc}
		\toprule
		Model & N-IA@1  & N-IA@3 & N-IA@5 & N-IA@10\\
		\midrule
		None(Random) & $0.4441^{**}$ & $0.4463^{**}$ & $0.4496^{**}$ & $0.4694^{**}$   \\
		Text Only & $0.5637^{*}$ & $0.5707^{**}$ & $0.5790^{**}$ & $0.6147^{**}$ \\
		Charge Only & 0.6340 & $0.6587^{*}$ & $0.6877^{*}$ & $0.7620^{**}$ \\
		\midrule
		DLRM & 0.7199 & 0.7389 & 0.7753 & 0.8747 \\
		\bottomrule
	\end{tabular}
	
	\label{ablation_NDCG}
\end{table}


\section{Conclusion and Future work}

In this paper, we study the need of search diversity in legal case retrieval.
We start from conducted a lab study and create the first legal case retrieval dataset with human annotation on intent-level document relevance.
By analyzing and modeling the relationships between different charges in queries and documents, we further propose a Diversified Legal case Retrieval Model that outperforms the state-of-the-art search diversification algorithms on legal case retrieval

This study is an initial study demonstrating the potential of search diversification in legal retrieval.
In future, we are interested in further analyzing the need of search diversity and how to incorporate different domain knowledge to improve legal retrieval models.
Also, the experiment evaluation in this paper is conducted with standard diversity metrics that are not tailored for legal retrieval.
How to develop effective evaluation and satisfaction prediction methods for legal case retrieval is also an important problem that we want to work in in the future.

\newpage


\bibliographystyle{ACM-Reference-Format}
\bibliography{sample-base}

\appendix

\end{document}